\documentstyle [12pt]{article}
\begin{document}

\hfill {WM-97-108}

\hfill {June, 1997}

\vskip 1in   \baselineskip 24pt
{

   \bigskip \centerline{\bf $K_L\rightarrow \pi^o\nu\bar{\nu}$ in the Aspon
Model} }
 \vskip .8in
\centerline{Arjun Berera and
Thomas W. 
Kephart}

\bigskip
\centerline{\it Department of Physics and Astronomy, Vanderbilt University,
Nashville, TN  37235, USA}
\bigskip
 
\centerline{Marc
Sher } 
\bigskip
\centerline {\it Physics Department, College of William and
Mary, Williamsburg, VA 23187, USA}

\vskip 1in
 
{\narrower\narrower   An attractive
alternative to the standard model of CP violation is the aspon model.  In this
model, CP is spontaneously broken, automatically solving the strong CP problem. 
Recently, it has been shown that CP violation in the B system is much smaller in
the aspon model than in the standard model.  
Here we provide a complementary study by considering $K_L\rightarrow
\pi^o\nu\bar{\nu}$, which is almost entirely CP violating and free of hadronic
uncertainties, and show that the rate in the aspon model is many orders of
magnitude below the standard model rate.  Observation of $K_L\rightarrow
\pi^o\nu\bar{\nu}$ would rule out the aspon model.}

\newpage
A promising alternative to the standard model explanation of CP
violation is the aspon model\cite{fk,fn,fnkw}.  In this model, CP is spontaneously
broken, and the strong CP problem is automatically resolved.  Recently, Frampton
and Glashow\cite{fg}, following work of Ackley, et al.\cite{ackley}, showed that
a consequence of the aspon model is that  CP violation
in the B-sector will be significantly suppressed, leading to a clear distinction
between this model and the standard model.  In this Brief Report, we point out
another important distinction between the two models:  the rate for
$K_L\rightarrow
\pi^o\nu\bar{\nu}$.

$K_L\rightarrow \pi^o\nu\bar{\nu}$ has been regarded as a ``gold-plated" test
of CP violation\cite{buras}.  It is almost entirely CP-violating, and the
calculation of the decay rate is free of hadronic uncertainties.  Although the
rate in the standard model is small, experiments are currently being planned
at Brookhaven National Laboratory and the Thomas Jefferson National Accelerator
Facility\cite{plans} to detect the decay, and measure the rate to
$10\%$ accuracy.  The calcuation of the rate involves determining the effective
flavor-changing
$s-d-Z$ vertex.  In the standard model\cite{inami}, the loop involves a $W$ and a
top quark, and the diagram is thus proportional to the product of CKM
elements: $V_{ts}V_{td}*$.  Since the decay is CP-violating, the effective vertex
is proportional to the imaginary part of this quantity. (Box diagrams are
proportional to lepton masses and are thus much smaller.)

In the aspon model, an additional $U(1)$ gauge group is added to the standard
model.  The standard model particles are singlets under this group.  One then
usually
adds a vectorlike 
doublet
of quarks, $U$ and $D$, and two 
complex
Higgs scalars, $\chi_1$
and $\chi_2$, which are 
charged
under the $U(1)$.  When the $\chi_i$ get
vacuum expectation values, the $U(1)$ is broken; the resulting massive gauge
boson is called the aspon.  Since, in general, the vacuum values of the
$\chi_i$ will have a nonzero relative phase, CP will be spontaneously broken by
this phase.  Since the Lagrangian is CP-invariant, there is no strong CP
problem and the CKM elements are real.  Thus, the standard model contribution
to $K_L\rightarrow\pi^o\nu\bar{\nu}$ will vanish.

There will be a contribution to the effective $s-d-Z$ vertex in the aspon model
from a loop involving  the $\chi$ fields and the vectorlike $D$-quark.
The couplings to the light quarks are given by
\begin{equation}
{\cal L}=(h^1_k\chi_1+h^2_k\chi_2)(\bar{d}^k_LD_R)+{\rm H.c.}
\end{equation}
The $h^i$ Yukawa couplings are real, by the CP invariance of the Lagrangian.
When the $\chi$ fields get vacuum expectation values, this term causes mixing
between the $D$ and $d^k$ fields. The mixing will, in general, be complex due
to the complex vacuum expectation values.
The mixing can be parametrized by the quantity $x_k$, which is
defined to be
$h^1_k{\langle\chi_1\rangle\over M}+h^2_k{\langle\chi_2\rangle\over M}$, where
$M$ is the $D$-quark mass.  Frampton and Glashow\cite{fg} show that a plausible
range for the $x_i^2$ is between $3\times 10^{-5}$ and $10^{-3}$.

So, how big is this loop in the aspon model?  The difference between this case
and the standard model case is that the couplings $g^2V_{ts}V_{td}*$ are
replaced by $h_2h_1\sim x_2x_1(M^2/\langle\chi\rangle^2)$   There is a
dependence on the masses in the loops (the $W$ and top in the standard model
and the $\chi$ and $D$ in the aspon model), of course; this dependence gives
a factor which is
$O(1)$ (but could vary by an order of magnitude or so).  Given that
$g^2V_{ts}V_{td}*$ is
$O(10^{-4})$, we see that, given Frampton and Glashow's plausible range for the
$x_i$, the two models have similar contributions to the effective $s-d-Z$
vertex.

However, it is crucial to note that the contribution to the 
$K_L\rightarrow \pi^o\nu\bar{\nu}$ decay depends on the {\it imaginary} part of
the vertex.  For the standard model, the imaginary part is still $O(10^{-4})$. 
In the aspon model, however, the $h_i$, as well as the CKM elements, are real. 
To get an imaginary contribution, one must either go to higher loops (causing a
suppression of roughly two orders of magnitude), or consider the loop with an
internal strange or down quark.  A $s-s-\chi$ or $s-d-\chi$ vertex can arise
from mixing between the $D$ with an $s$ or $d$.  However, this mixing will
be proportional to $x_2x_1$, and this gives an additional factor of $x_2x_1 <
10^{-3}$, suppressing the result by at least three orders of
magnitude.\cite{note}

Just as it has been shown that CP violation in the B-system in the
aspon model is severely suppressed, we have shown here that $K_L\rightarrow
\pi^o\nu\bar{\nu}$ is also suppressed.  Discovery of such a decay would rule
out the aspon model.

\def\prd#1#2#3{{\rm Phys.~Rev.~}{\bf D#1}, #2 (19#3)}
\def\plb#1#2#3{{\rm Phys.~Lett.~}{\bf B#1}, #2 (19#3)}
\def\npb#1#2#3{{\rm Nucl.~Phys.~}{\bf B#1}, #2 (19#3)}
\def\prl#1#2#3{{\rm Phys.~Rev.~Lett.~}{\bf #1}, #2 (19#3)}

\bibliographystyle{unsrt}

\begin{thebibliography}{99}
\bibitem{fk}
P.H. Frampton and T.W. Kephart, \prl{66}{1666}{91}.
\bibitem{fn}
P.H. Frampton and D. Ng, \prd{43}{3034}{91}.
\bibitem{fnkw}
P.H. Frampton, D. Ng, T.W. Kephart, and T. J. Weiler \prl{68}{2129}{1992}.
\bibitem{fg}
P.H. Frampton and S.L. Glashow, \prd{55}{1691}{97}
\bibitem{ackley}
A.W. Ackley, P.H. Frampton, B. Kayser and C.B. Leung, \prd{50}{3560}{94}.
\bibitem{buras}
A.J. Buras, \plb{333}{476}{94}; A.J. Buras, in {\it Proceedings of the
International Conference on High Energy Physics}, Glasgow, Scotland, 1994,
edited by P.J. Bussey and I.G. Knowles (IOP, London, 1995).
\bibitem{plans}
L. Littenberg and M. Ito (private communication).
\bibitem{inami}
T. Inami and C.S. Lim, {\rm Prog. Theor. Phys.~}{\bf 65}, 297 (1981).
\bibitem{note}
Note that the related decay
$K^+\rightarrow \pi^+\nu\bar{\nu}$ is not CP-violating, and thus the standard
model contribution is not eliminated by the reality of the CKM matrix. One
expects that decay to have a contribution from the aspon model which is of
similar order of magnitude.  Since the calculation of this process is much less
clean, it would be difficult to make definitive statements about the aspon model
from  detection of this decay mode.
\end{thebibliography}

\end{document}